\documentclass[10pt,superscriptaddress,twocolumn,amsmath,amssymb,aps,pra,showpacs]{revtex4-1}
\usepackage{mathrsfs}
\usepackage{graphicx}
\usepackage{dcolumn}
\usepackage{bm}
\usepackage[title,titletoc,header]{appendix}

\usepackage{amssymb}
\usepackage{amsmath}
\usepackage{mathrsfs}
\usepackage{amsmath}
\usepackage{subfigure}
\usepackage{float}
\usepackage[title,titletoc,header]{appendix}
\usepackage{graphicx}

\newcommand{\be}{\begin{equation}}
\newcommand{\ee}{\end{equation}}
\newcommand{\bea}{\begin{eqnarray}}
\newcommand{\eea}{\end{eqnarray}}

\begin{document}
\title{Proposed realization of critical regions  in a one-dimensional flat band lattice with a quasi-periodic potential }

\author{Yi-Cai Zhang}
\affiliation{School of Physics and Materials Science, Guangzhou University, Guangzhou 510006, China}




\date{\today}
\begin{abstract}
In the previous work, the concept of critical region in a generalized Aubry-Andr\'{e}  model (Ganeshan-Pixley-Das Sarma's model) has been set up.
In this work we propose that the critical region can be realized in a one-dimensional flat band lattice system  with a quasi-periodic potential.
 It is found that the above flat band lattice model can be reduced into an effective Ganeshan-Pixley-Das Sarma's model
where the effective parameter $\alpha=V_0/(2E)$ with potential strength $V_0$ and eigenenergy $E$. It is shown that  there are very rich physics in this model.
Depending on $|\alpha|<1$ or $|\alpha|\geq1$, the effective quasi-periodic potential would be bounded or unbounded. For these two cases, the Lyapunov exponent [$\gamma(E)$], mobility edges ($E_c$) and critical indices ($\nu$) of localized length are obtained exactly.
  In addition, several localized state regions, extended state regions and critical regions would appear in the parameter $V_0-E$ plane.
For a given potential strength $V_0$, the localized-extended and localized-critical transitions can co-exist.
 Furthermore, we find the critical index of localized length $\xi(E)=1/\gamma(E)$ is $\nu=1$  near localized-extended transitions  and $\nu=1/2$ near the localized-critical transitions.
 Near the transition point between the bound ($|\alpha|<1$) and unbounded ($|\alpha|\geq1$) cases, i.e, $|\alpha|=|V_0/(2E)|= 1$, the derivative of Lypunov exponent of localized states with respect to energy is discontinuous.   The localized states in bounded and unbounded cases can be distinguished from each other by Avila's acceleration.  At the end, we find that near the transition point,  there also exist critical-extended transitions in the phase diagram.

\end{abstract}

\maketitle
\section{Introduction}
A lot of  novel physics, for example, existences of localized flat band states \cite{Sutherland1986,Vidal1998,Mukherjee}, ferro-magnetism transition \cite{Mielke1999,Zhang2010}, super-Klein tunneling \cite{Shen2010,Urban2011,Fang2016,Ocampo2017}, preformed pairs \cite{Tovmasyan2018}, strange metal \cite{Volovik2019}, high $T_c$ superconductivity/superfluidity \cite{Peotta2015,Hazra2019,Cao2018,Wuyurong2021,Kopnin2011,Julku2020,Iglovikov2014,Julku2016,Liang2017,Iskin2019,Wu2021}, etc., can  appear in a flat band system.
Due to infinitely large density of states of flat band, a short-ranged potential can result in an infinite number of  bound states, even a hydrogen atom-like  energy spectrum, i.e., $E_n\propto1/n^2,n=1,2,3,...$ \cite{Zhangyicai2021}.
Furthermore,
a long ranged Coulomb potential can destroy completely the flat band \cite{Gorbar2019,Pottelberge2020}.
In addition, a long-ranged Coulomb potential can result in wave function collapse \cite{Han2019,Zhangyicai20212} and a $1/n$ energy spectrum \cite{Zhangyicai20213}, even the bound states in a continuous spectrum (BIC)  \cite{Zhangyicai20214}.

In the past decades, Anderson localization and the existences of mobility edges in one-dimensional lattice model with quasi-periodic potentials have attracted a great interests \cite{Sarma1988,Sarma1990,Tang2021,Sil2008,Biddle2010,Liu2017,Longhi2019,Duthie2021,Liu2021,An2021}.
A famous example where the localized-extended transition can occur is the Aubry-Andr\'{e } lattice model (AA model) \cite{Aubry1980},
i.e.,
\begin{align}\label{10}
t[\psi(n+1)+\psi(n-1)]+2\lambda \cos(2\pi\beta n+\phi)\psi(n)=E\psi(n).
\end{align}
where $t$ is hopping, $n\in Z$ is lattice site index, $2\lambda$ describes the quasi-periodic potential strength, $\beta$ is an irrational number, $\phi$ is a phase.
When the quasi-periodic  potential weak, all the eigenstates are extended states. While when the potential strength is sufficiently large, all the eigenstates become localized states.
In this model, there are no mobility edges. The non-existences of mobility edges originate from the Aubry-Andr\`{e} self-duality of this model.
However, the breaking of the self-duality would result in the appearance of mobility edges \cite{Delyon1984,Sarma1988,Sarma1990,Izrailev1999,Boers2007, Sil2008,Liu2017,Li2017,Luschen2018,Tang2021}.

A generalized Aubry-Andre model (GAA model) which can have exact mobility edges has been proposed by Ganeshan, Pixley and Das Sarma \cite{Ganeshan2015}.
 The GAA model is
 \begin{align}\label{2}
t[\psi(n+1)+\psi(n-1)]+\frac{2\lambda \cos(2\pi\beta n+\phi)}{1-\alpha \cos(2\pi\beta n+\phi)}\psi(n)=E\psi(n).
\end{align}
In comparison with the AA model, there is an additional real parameter $\alpha$ in the denominator of quasi-periodic potential $\frac{2\lambda cos(2\pi\beta n+\phi)}{1-\alpha cos(2\pi\beta n+\phi)}$.
Interestingly, the mobility edges can be exactly obtained with a generalized self-duality transformation.
Recently, a so-called mosaic lattice model has been proposed which also has mobility edges \cite{Wangyucheng2020}. The mobility edges can be also exactly obtained with Avila's theory \cite{YONGJIAN2,Liu2021,Avila2015}.

In the most of the previous studies of GAA model, $\alpha$ is mainly limited to $|\alpha|<1$ due to concern of possible appearances of divergences of periodic potential [see Eq.(2)]. Then, the quasi-periodic potential is bounded. Very recently, the localization problem for the unbounded ($|\alpha|\geq1$) case has been investigated by the present author and the co-author \cite{Zhangyicai2022}. It is found that when $|\alpha|\geq1$, the quasi-periodic potential and energy spectrum would be unbounded.
In addition, there exist a critical region which consists of critical states in a parameter plane. When the energy approaches the localized-critical transition point (mobility edge), the critical index of localized length $\nu=1/2$, which is different from $\nu=1$ of bounded $|\alpha|<1$ case. In addition, it is found that  Avila's acceleration for unbound case is also quantized. The systems with different $E$ can be classified by the Lyapunov exponent and Avila's acceleration.

Due to the divergences of unbounded quasi-periodic potential in original GAA model, a natural question arises: is it possible to realize the critical region without any divergences in energy spectrum?
In addition, for a given quasi-periodic potential strength, it is desirable that if one can realize the above bounded and unbounded quasi-periodic potentials in a simple model. In addition, one may continue to ask a question: can the localized-extended transition and localized-critical transition co-exist?

In this work, we find that in the presence of quasi-periodic potential, the above objectives can be realized in  a simple flat band lattice model. This lattice model can be reduced into an effective GAA model. Especially, the unbounded case can be realized without any divergences in the energy spectrum.
It is found that the localized, extended and critical states can co-exist in the one-dimensional flat band lattice model with a quasi-periodic potential.
The two distinct transitions, i.e.,  localized-extended and localized-critical transitions also appear in the potential strength $V_0$-$E$ (energy) plane.

The work is organized as follows. In Sec.\textbf{II}, a model Hamiltonian and its three energy bands are given.  Next, we  investigate the localization question of type III quasi-periodic potential in Sec.\textbf{III}.
 At the end, a summary is given in Sec.\textbf{IV}.

\begin{figure}
\begin{center}
\includegraphics[width=1.0\columnwidth]{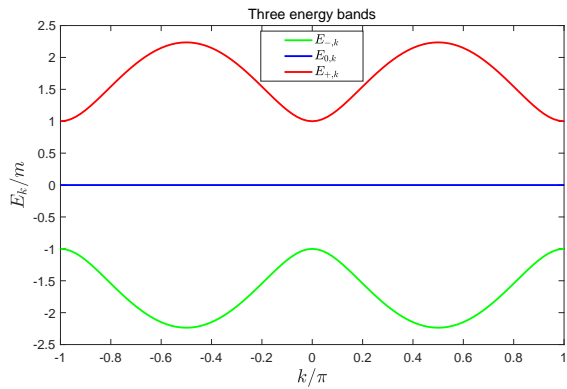}
\end{center}
\caption{ The three energy bands of free particle Hamiltonian $H_0$. There exists a flat (middle) band ($E_{0,k}$) in between the upper band ($E_{+,k}$) and the lower band ($E_{-,k}$).}
\label{schematic}
\end{figure}

\section{A tightly binding Hamiltonian  with a flat band}
In this work, we consider a tightly binding lattice Hamiltonian with three sublattices A, B, and C, i.e.,
 \begin{align}
&H=H_0+V_p\notag\\
&H_0=-\frac{it}{\sqrt{2}}\sum_{n\in Z}[a^{\dag}_{n-1}b_{n}+b^{\dag}_{n-1}a_{n}+b^{\dag}_{n-1}c_{n}+c^{\dag}_{n-1}b_{n}]+\emph{h.c.}\notag\\
&+m\sum_{n\in Z}[a^{\dag}_{n}a_{n}-c^{\dag}_{n}c_{n}],
    	\label{hamiltonian}
\end{align}
where $V_p$ is potential energy, integer $n$ is the lattice site index, $H_0$ is the free-particle Hamiltonian, $t>0$ is hopping parameter, and $m>0$ is energy gap parameter.
$a(b/c)_{n}$ are the annihilation operators for states at sublattices $A(B/C)$, respectively.

 When potential $V_p=0$, applying a Fourier transform, the free particle Hamiltonian $H_0$ can be written as
   \begin{align}
&H_0=\sqrt{2}t\sum_{-\pi\leq k\leq \pi}\sin(ka)[a^{\dag}_{k}b_{k}+b^{\dag}_{k}a_{k}+b^{\dag}_{k}c_{k}+c^{\dag}_{k}b_{k}]\notag\\
&+m\sum_{-\pi\leq k\leq \pi}[a^{\dag}_{k}a_{k}-c^{\dag}_{k}c_{k}],
    	\label{hamiltonian}
\end{align}
where $a$ is lattice constant. In the whole manuscript, we would set $a=1$ for simplifications.

Furthermore, in the above Hamiltonian $H_0$, we can identify the above three sublattices $A,B,C$ as three spin components $1,2,3$. In the spin basis $|1,2,3\rangle$,
   the three eigenstates  and the
 eigenenergies are
\begin{align}
&\langle x|-, k\rangle=\frac{e^{ikx}}{2\sqrt{4t^2\sin^2(k)+m^2}}\left(\begin{array}{ccc}
\sqrt{4t^2\sin^2(k)+m^2}-m\\
 -2\sqrt{2}t\sin(k)\\
\sqrt{4t^2\sin^2(k)+m^2}+m
  \end{array}\right),\notag\\
  &E_{-, k}=-\sqrt{4t^2\sin^2(k)+m^2};\notag\\
  &\langle x|0, k\rangle=\frac{e^{ikx}}{\sqrt{4t^2\sin^2(k)+m^2}}\left(\begin{array}{ccc}
-\sqrt{2}t\sin(k)\\
 m\\
\sqrt{2}t\sin(k)
  \end{array}\right),\notag\\
  &E_{0, k}=0;\notag\\
  &\langle x|+, k\rangle=\frac{e^{ikx}}{2\sqrt{4t^2\sin^2(k)+m^2}}\left(\begin{array}{ccc}
\sqrt{4t^2\sin^2(k)+m^2}+m\\
 -2\sqrt{2}t\sin(k)\\
\sqrt{4t^2\sin^2(k)+m^2}-m
  \end{array}\right),\notag\\
  &E_{+, k}=\sqrt{4t^2\sin^2(k)+m^2},
\end{align}
where $|-, 0, +; k\rangle$ denote the the eigenstates of lower, middle (flat) and upper bands, respectively and $E_{-, 0, +; k}$ represent the corresponding three energy bands.
It is found that a flat band with zero energy ($E_{0, k}=0$) appears in between upper and lower bands (see Fig.1).

It is noted that near the momentum $k=0$, the above three band Hamiltonian would be transformed into a continuous spin-1 Dirac model with a flat band.
The bound state problems of the continuous version for above Hamiltonian with various types of  potentials have been investigated by Zhang and Zhu \cite{Zhangyicai2021,Zhangyicai20212,Zhangyicai20213,Zhangyicai20214}.

\begin{figure}
\begin{center}
\includegraphics[width=1.0\columnwidth]{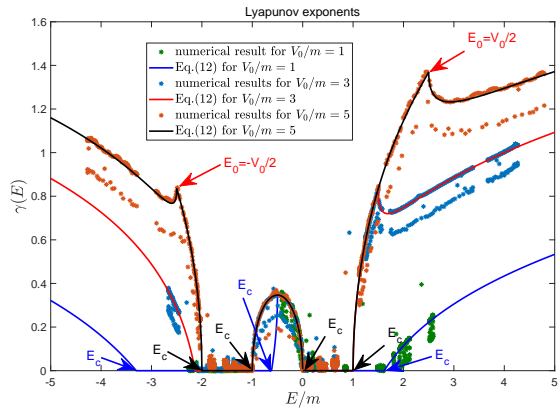}
\end{center}
\caption{Lyapunov exponents for potential strength $V_0/m=1,3,5$.   The discrete points are the numerical results for eigenenergies. The solid lines are given by Eq.(\ref{Coul}). The mobility edges for  $V_0/m=1,5$  are indicated by blue arrows and black  arrows, respectively.  Near mobility edges of the localized-extended transition (e.g., $E_c\simeq-3.30m$,-0.62m, and $1.62m$ for $V_0/m=m$ and $|\alpha|<1$), the Lyapunov exponent $\gamma(E)\propto |E-E_c|$ approaches zero. The critical index of the localized length $\nu$ is $1$ for $|\alpha|<1$.   While $E$ is near the localized-critical  transition (e.g., $E_c=-2m$, $-m$, $0$ and $m$ for $V_0/m=5$ and $|\alpha|>1$), the Lyapunov exponent $\gamma(E)\propto |E-E_c|^{1/2}$ (as $E\rightarrow E_c$), and the critical index of the localized length $\nu=1/2$.
The transition points ($|\alpha|=1$) $E=E_0\equiv \pm V_0/2=\pm2.5m$ between bounded and unbounded cases  for $V_0=5m$ are also indicated by red arrows. In the whole manuscript,  we take $t=m$ and phase $\phi=0$. }
\label{schematic}
\end{figure}

\begin{figure}
\begin{center}
\includegraphics[width=1.0\columnwidth]{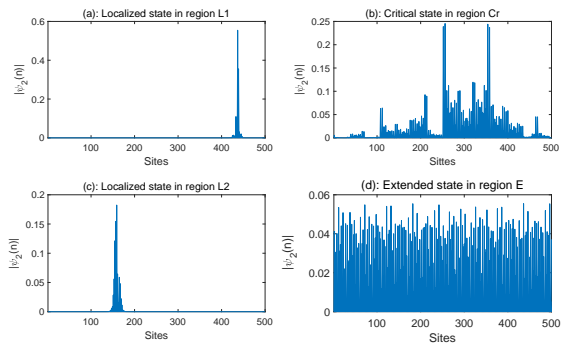}
\end{center}
\caption{ Several typical wave functions for extended, localized, and critical states.}
\label{schematic}
\end{figure}

\section{ Localized-extended and localized-critical transitions in  a quasi-periodic potential of type III }
In the following, we assume the potential energy $V_p$ has following form in spin basis $|1,2,3\rangle$, namely,
\begin{align}\label{3}
&V_p=V_{11}(n)\bigotimes|1\rangle\langle1|
=\left[\begin{array}{ccc}
V_{11}(n) &0  & 0\\
0&0& 0\\
0 &0 & 0
  \end{array}\right],
\end{align}
where quasi-periodic potential
\begin{align}
V_{11}(n)=V_0\cos(2\pi\beta n+\phi)\nonumber
\end{align}
with potential strength $V_0$.
In the whole manuscript, we would refer such a kind of potential as potential of type III \cite{Zhangyicai20213}.
The bound state problems with potential of type I and II have been investigated
by the present author \cite{Zhangyicai2021,Zhangyicai20212}.

The Schr\"{o}dinger equation ($H\psi=E\psi$) can be written in terms of three component wave functions, i.e.,
\begin{align}\label{9}
&it[\psi_{2}(n+1)-\psi_{2}(n-1)]/\sqrt{2}=[E-m-V_{11}(n)]\psi_{1}(n),\notag\\
&\frac{it}{\sqrt{2}}[\psi_{1}(n+1)-\psi_{1}(n-1)+\psi_{3}(n+1)-\psi_{3}(n-1)]=E\psi_{2}(n),\notag\\
&i[\psi_{2}(n+1)-\psi_{2}(n-1)]/\sqrt{2}=[E-m]\psi_{3}(n).
\end{align}
Adopting a similar procedure as Ref. \cite{Zhangyicai20213}, using Eq.(\ref{9}) to eliminate wave functions of 2-th and 3-th components, we get an effective equation for $\psi_1(n)$
\begin{align}\label{10}
&t^2[\frac{E-V_{11}(n+2)/2}{E+m}\psi_{1}(n+2)-2\frac{E-V_{11}(n)/2}{E+m}\psi_{1}(n)\notag\\
&+\frac{E-V_{11}(n-2)/2}{E+m}\psi_{1}(n-2)]=-E[E-m-V_{11}(n)]\psi_{1}(n).
\end{align}
Further we introduce an auxiliary
wave function $\psi(n)\equiv\frac{E-V_{11}(n)/2}{E+m}\psi_{1}(n)$, and effective hopping $\tilde{t}$, effective total energy $\tilde{E}$, effective potential strength $\lambda$ and effective parameter $\alpha$, i.e.,
\begin{align}\label{10}
&\tilde{t}\equiv t^2,\notag\\
&\tilde{E}\equiv-E^2+m^2+2\tilde{t}=-E^2+m^2+2t^2,\notag\\
&\lambda\equiv-\frac{V_0(E+m)^2}{4E},\notag\\
&\alpha\equiv\frac{V_0}{2E},
\end{align}
 then we get an equation for $\psi(n)$
 \begin{align}\label{11}
&\tilde{t}[\psi(n+2)+\psi(n-2)]+\frac{2\lambda \cos(2\pi\beta n+\phi)}{1-\alpha \cos(2\pi\beta n+\phi)}\psi(n)=\tilde{E}\psi(n).
\end{align}
This is an effective generalized Aubry-Andr\'{e} model  whose effective lattice constant is two times original lattice constant, i.e., $2a=2$.

We should remark that when the effective parameter $|\alpha|\geq1$, the above effective GAA model Eq.(\ref{11}) has an unbounded quasi-periodic potential superficially. However, differently from the original  Ganeshan-Pixley-Das Sarma's GAA model \cite{Zhangyicai2022}, here the energy spectrum is still bounded.
This is because that $|\alpha|\geq1$ implies that
\begin{align}
|\alpha|=|\frac{V_0}{2E}|\geq1\Rightarrow  |E|\leq \frac{|V_0|}{2}.
\end{align}
It shows that for a given potential strength $V_0$, the energy $E$ is always bounded.
This can be also understood as follows. In the original three-component lattice model, i.e.,  Eq.(\ref{9}), the quasi-periodic potential $V_{11}$ has no any singularities and divergences, then
the energy spectrum should be bounded.

In following text, we will show that although the energy spectrum is bounded, all the other interesting physical phenomena of unbounded quasi-periodic potentials, for example, the existence of critical region, localized-critical transition and  a different critical index can still appear in the above effective GAA model Eq.(\ref{11}). Then, in the whole manuscript, we still call  $|\alpha|=|\frac{V_0}{2E}|\geq1$ as unbounded case.

The localized properties of eigenstates can be characterized by Lyapunov exponent. With the Avila's theory \cite{Liu2021,Avila2015,YONGJIAN2},  Lyapunov exponents have been exactly obtained for both bounded ($|\alpha|<1$) and unbounded ($|\alpha|\geq1$) quasi-periodic potentials in Ref.\cite{Zhangyicai2022}, i.e.,
\begin{align}\label{Coul}
&\gamma(E)=
\left\{\begin{array}{cccc}
\frac{1}{2}Max\{0,\log(\frac{|P|+\sqrt{P^2-4\alpha^2}}{2|1+\sqrt{1-\alpha^2}|})\},   \ |\alpha|<1 \ \& \ P^2>4\alpha^2\\
0, \ \ \ \ \ \ \ \ \ \ \ \ \ \ \ \ \ \ \ \ \ \ \ \ \ \ \ \ \ \ \ \ \ |\alpha|<1 \ \& \ P^2<4\alpha^2\\
\frac{1}{2}\log(\frac{|P|+\sqrt{P^2-4\alpha^2}}{2|\alpha|}),\ \ \ \ \ \ \ \ \ \ \ \ |\alpha|\geq1 \ \& \ P^2>4\alpha^2\\
0.\ \ \ \ \ \ \ \ \ \ \ \ \ \ \ \ \ \ \ \ \ \ \ \ \ \ \ \ \ \ \ \ \ \ |\alpha|\geq1 \ \& \ P^2<4\alpha^2.
  \end{array}\right.
\end{align}
where
\begin{align}\label{H0}
P=\frac{\alpha \tilde{E}+2\lambda}{\tilde{t}}=\frac{\alpha \tilde{E}+2\lambda}{t^2}.
\end{align}
It is noted that in comparison with Eq.(26) of Ref.\cite{Zhangyicai2022}, due to two times original lattice constant here, there is an extra factor $1/2$ in Eq.(\ref{Coul}).

In order to characterize the properties of eigenstates, we also solve  Eq.(\ref{9}) numerically.  To be specific, we take total lattice site number $N= 500$  and a $3N\times3N$ matrix can be established with open boundary conditions at  two end sites. Then, we diagonalize it to get the 1500 eigenenergies and eigenstates. The results are reported in Figs.2, 3, 4 and 5.

In addition, with Eq.(\ref{11}), the Lyapunov exponent can be calculated numerically with transfer matrix method \cite{Zhangyicai2022}, i.e.,
\begin{align}\label{V}
&\gamma(E)=\lim_{L \rightarrow \infty }\frac{\log(|\Psi(2L)|/|\Psi(0)|)}{2L}\notag\\
&=\lim_{L\rightarrow \infty}\frac{\log(|T(2L)T(2L-2)...T(4)T(2)\Psi(0)|/|\Psi(0)|)}{2L}
\end{align}
where $L$ is a positive integer, transfer matrix
\begin{align}
T(n)\equiv\left[\begin{array}{ccc}
\frac{\tilde{E}}{\tilde{t}}-\frac{2\lambda}{\tilde{t}} \cos(2\pi\beta n+\phi) &-1  \\
1&0\\
  \end{array}\right],
\end{align}
\begin{align}
\Psi(n)\equiv\left[\begin{array}{ccc}
\psi(n+2)  \\
\psi(n)\\
  \end{array}\right],
\end{align}
and
\begin{align}
|\Psi(n)|=\sqrt{|\psi(n+2)|^2+|\psi(n)|^2}.
\end{align}
 To be specific, taking $t/m=1$, $V_0/m=1,3,5$, we calculate the Lyapunov exponents numerically for all the eigenenergies [see the  sets of discrete points in  Fig.2].
 In our numerical calculation, we take $L=125$, phase $\phi=0$, $\psi(0)=0$ and $\psi(2)=1$ in Eq.(\ref{V}).
  The solid lines of Fig.2 are given by Eq.(\ref{Coul}) with the same parameters. It is shown that most of all discrete points fall onto the solid lines.

  However, we also note that there are some discrete points of localized states which are not on the solid lines. This is because these localized wave functions  are too near the left-hand boundary of system.

It is shown that, depending on $|\alpha|<1$  or $|\alpha|\geq1$,  there exist two types of localized-delocalized transitions in the GAA model \cite{Zhangyicai2022}.
If  $|\alpha|<1$, the effective quasi-periodic potential in Eq.(\ref{11}) is bounded, there are  mobility edges which separate the localized states from the extended states, i.e.,  localized-extended transitions.
While when $|\alpha|\geq1$, the quasi-periodic potential is unbounded, there would exist critical regions which consists of critical states. In such a case, the mobility edges separate the localized states from critical states (see Ref.\cite{Zhangyicai2022} and Fig.4).

When $|\alpha|<1$, by Eqs.(\ref{Coul}) and (\ref{10}), the mobility edges are given by
\begin{align}
&\gamma(E=E_c)=0\rightarrow |P|=2, \rightarrow \notag\\
&|\frac{V_0}{E_c}-\frac{V_0(m+E_c)}{t^2}|=2.
\end{align}
The extended state regions are given by
\begin{align}
|P|<2 \rightarrow |\frac{V_0}{E}-\frac{V_0(m+E)}{t^2}|<2.
\end{align}

When $|\alpha|\geq1$, by Eqs.(\ref{Coul}) and (\ref{10}), the mobility edges are given by
\begin{align}
&\gamma(E=E_c)=0\rightarrow |P|=2|\alpha|,\notag\\
& \rightarrow |\frac{t^2-(m+E_c)E_c}{t^2}|=1.
\end{align}
The critical regions are defined by
\begin{align}
&|P|<2|\alpha| \rightarrow |\frac{t^2-(m+E)E}{t^2}|<1.
\end{align}

When the energy of localized state approaches  localized-delocalized (extended and critical) transition point (mobility edge), the Lyapunov exponent goes to zero according to the law of
\begin{align}
\gamma(E)\propto |E-E_c|^\nu\rightarrow0,
\end{align}
where $\nu>0$ is critical index of localized length of localized states.
Consequently, the localized length
\begin{align}
\xi(E)\equiv1/\gamma(E)\propto |E-E_c|^{-\nu}\rightarrow\infty
\end{align}
becomes infinitely large. For a given potential strength $V_0$, near the mobility edges $E_c$ at which the localized-extended transition occurs, the localized length is \cite{Zhangyicai2022}
\begin{align}
\xi(E)=1/\gamma(E)\propto |E-E_c|^{-1},
\end{align}
while for the localized-critical transition,
\begin{align}
\xi(E)=1/\gamma(E)\propto |E-E_c|^{1/2}.
\end{align}
So the critical index  is $\nu=1$ for $|\alpha|<1$, and $\nu=1/2$ for $|\alpha|\geq1$ (see Fig.2).

  Several typical wave functions for the localized states, extended states and critical states are reported in Fig.3.
We can see that the localized states usually only occupy finite lattice sites, while the extended states occupy the whole lattice. The wave functions of critical states are composed of several disconnected patches.

The phase diagram in $V_0-E$ plane is reported in Fig.4. From Fig.4, we see that for a given $V_0$, the localized state regions (labeled with $L_1$ and $L_2$), extended state regions $E$ and critical regions $C_r$ can co-exist.

\begin{figure}
\begin{center}
\includegraphics[width=1.0\columnwidth]{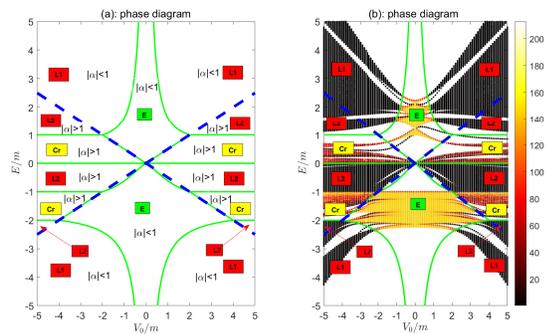}
\end{center}
\caption{ Phase diagram and standard deviations for quasi-periodic potential of type III. The extended state regions and the critical regions are labeled with $E$ and $C_r$, respectively. The two kinds of localized state regions are denoted with $L_1$ (for $|\alpha|<1$) and $L_2$ (for $|\alpha|\geq1$), respectively. (a): the boundaries between bounded and unbounded quasi-periodic potentials are defined by $|\alpha|=|\frac{V_0}{2E}|=1$ (the blue dashed lines).
 (b): standard deviations are represented with different colors.}
\label{schematic}
\end{figure}

\begin{figure}
\begin{center}
\includegraphics[width=1.0\columnwidth]{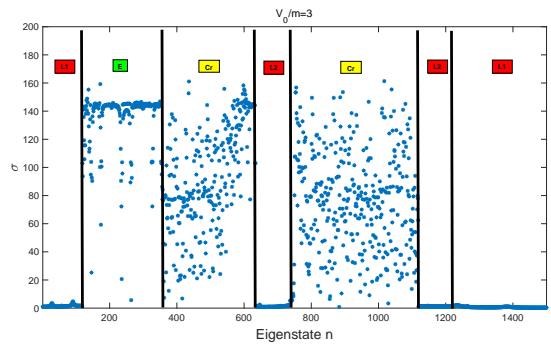}
\end{center}
\caption{ Standard deviations for  localized states, extended states and critical states. We take the potential strength $V_0/m=3$. The energy  $E$ of states increases as state index $n$ runs from $1$ to $1500$.}
\label{schematic}
\end{figure}

In order to distinguish the localized states from the extended states (and critical states), similarly as that in usual GAA model \cite{Zhangyicai2022},  here we also numerically calculate standard deviation of coordinates \cite{Boers2007}
\begin{align}\label{26}
&\sigma=\sqrt{\sum_{\sigma=1,2,3;n}(n-\bar{n})^2|\psi_{\sigma}(n)|^2},
\end{align}
where the average value of coordinate is
\begin{align}
\bar{n}=\sum_{\sigma=1,2,3;n}n|\psi_{\sigma}(n)|^2.
\end{align}
The standard deviation of coordinates describes the spatial extensions of wave functions.
When the states are localized, the standard deviations of coordinates are small. While for extended states, the standard deviations are much larger.
When the states are critical, their standard deviations are in between above two (see Fig.4).
In the comparison with localized and extended states, the critical states also have much larger fluctuations in the standard deviations (see Fig.5 and Ref.\cite{Zhangyicai2022}).

In addition, when $\alpha$ approaches the boundaries between  bounded and unbounded quasi-periodic potentials, i.e, $\alpha=\pm 1$ (see the blue dashed lines of Fig.4), there exist critical-extended transitions in the phase diagram.

Finally, for bounded $|\alpha|<1$ and unbounded $|\alpha|\geq1$ cases,   there are two kinds of localized state regions which are denoted by $L_1$ and $L_2$ in Fig.4.
Near the localized ($L_1$)-localized  ($L_2$) transitions, i.e, $|\alpha|=1$, we find that the derivative of Lyapunov exponents with respect energy $E$,  i.e., $\gamma'(E)\equiv\frac{d \gamma(E)}{dE}$ is discontinuous. For example, when $V_0=5m$, as the energy $E$ approaches the localized-localized transition point ($E_0=\pm V_0/2=\pm2.5m$), the derivative of Lyapunov exponent on the $L_2$ side is finite, while it diverges on the $L_1$ side, i.e, $\gamma'(E)\propto 1/\sqrt{|E-E_0|}\rightarrow \infty$ (see black line in Fig.2).


Furthermore, although both these two types of  localized states for bounded $|\alpha|<1$ and unbounded $|\alpha|\geq1$ cases have positive Lyapunov exponent i.e., $\gamma(E)>0$,
 they can  be still distinguished by Avila's acceleration $\omega(E)$ \cite{Zhangyicai2022}. For example,
when the localized state is in region $L_1$ which corresponds  the bounded $|\alpha|<1$ case, the Avila's acceleration $\omega(E)=1/2$.
While when the localized state is in region $L_2$ which corresponds  the unbounded $|\alpha|\geq1$ case, the Avila's acceleration $\omega(E)=0$, i.e.,
\begin{align}\label{21}
&\omega(E)=
\frac{1}{2}\times\left\{\begin{array}{c}
1,   \ for \ bound \ states \ of \ bounded \ case  \\
\ \ \ \ 0,  \ for \ bound \ states \ of \ unbounded \ case .
\end{array}\right.
\end{align}
It is noted that  due to two times original lattice constant here [see Eq.(\ref{11})], in comparison with the original GAA model (see Eqs. (45) and (47) of Ref.\cite{Zhangyicai2022}), there is also  an extra factor $1/2$ in Eq.(\ref{21}).

\section{summary}
In conclusion, we investigate the Anderson localization problem in  a one-dimensional flat band lattice model with a quasi-periodic potential.
 It is found that for type III potential, the localized states, extended states and critical regions can co-exist. With the variations of energy, there exist localized-extended and localized-critical transitions.
 For localized-extended transition, near the mobility edges, the Lypunov exponent goes to zero according to law of $|E-E_c|$. While for localized-critical transition, the Lyapunov exponent approaches zero by $|E-E_c|^{1/2}$. Consequently, for localized-extended transition, the critical index $\nu=1$. While for localized-critical transition, the critical index $\nu=1/2$.
In addition, near the transitions between bounded and unbounded quasi-periodic potentials, there exist localized ($L_1$)-localized ($L_2$) and critical-extended transitions in phase diagram.
Furthermore, when energy $E$ crosses the localized $L_1$-localized $L_2$ transitions, the derivative of Lyapunov exponent with respect to energy is discontinuous.
The localized states in $L_1$ and $L_2$ can be distinguished from each other by Avila's acceleration.

 In the presence of quasi-periodic potential of type II, the critical region would not appear (see \textbf{Appendix A}). There only exist localized-extended transitions and the critical index $\nu=1$.

Finally, in comparison with original Ganeshan-Pixley-Das Sarma's GAA model,  there are much richer physics in the flat band lattice model here.
Due to the divergences of quasi-periodic potential in the original GAA model, the unbounded quasi-periodic potential may only have theoretical interests  and its experimental realizations may be unrealistic.
Due to free of any divergences here, it is expected that the localization physics of the flat band lattice with quasi-periodic potential are much  more likely to be realized experimentally \cite{An2021}.

\appendix
\section{localized-extended transitions and mobility edges in  quasi-periodic potential of Type II}
In this appendix, for completeness, we would investigate the Anderson localization problem for quasi-periodic potential of Type II in the flat band lattice model.
We assume the potential energy $V_p$ has following form in spin basis $|1,2,3\rangle$, namely,
\begin{align}\label{A1}
&V_p=V_{22}(n)\bigotimes|2\rangle\langle2|
=\left[\begin{array}{ccc}
0 &0  & 0\\
0&V_{22}(n)& 0\\
0 &0 & 0
  \end{array}\right].
\end{align}
The potential only appears in the basis element $|2\rangle$ (or sublattice B).
In the whole manuscript, we would refer such a kind of potential as potential of type II \cite{Zhangyicai2021}.

Furthermore, we assume  $V_{22}$ is  also a quasi-periodical potential, i.e.,
\begin{align}\label{A2}
V_{22}(n)=V_0\cos(2\pi\beta n+\phi),
\end{align}
where $V_0$ is the potential strength, irrational number $\beta$ determines the quasi-periodicity, and real number $\phi$ is a phase.

The Schr\"{o}dinger equation ($H\psi=E\psi$) can be written as
\begin{align}\label{A3}
&it[\psi_{2}(n+1)-\psi_{2}(n-1)]/\sqrt{2}=[E-m]\psi_{1}(n),\notag\\
&it[\psi_{1}(n+1-\psi_{1}(n-1)+\psi_{3}(n+1)-\psi_{3}(n-1)]/\sqrt{2}\notag\\
&=[E-V_{22}(n)]\psi_{2}(n),\notag\\
&i[\psi_{2}(n+1)-\psi_{2}(n-1)]/\sqrt{2}=[E-m]\psi_{3}(n).
\end{align}
Adopting a similar procedure as Ref. \cite{Zhangyicai2021} to eliminate wave functions for 1-th and 3-th components, we get an effective discrete Schr\"{o}dinger  equation for $\psi_2$
\begin{align}\label{A4}
&-t^2[\psi_{2}(n+2)-2\psi_{2}(n)+\psi_{2}(n-2)]\notag\\
&+\frac{(E^2-m^2)V_{22}(n)}{E}\psi_{2}(n)=[E^2-m^2]\psi_{2}(n).
\end{align}
Further introducing an  effective hopping $\tilde{t}$, effective energy $\tilde{E}$ and effective potential strength $\lambda$
\begin{align}\label{A5}
&\tilde{t}\equiv t^2,\notag\\
&\tilde{E}\equiv -E^2+m^2+2t^2,\notag\\
&\lambda\equiv V_{0}(-E^2+m^2)/(2E),
\end{align}
Eq.(\ref{A4}) becomes the well-known  Aubry-Andr\'{e} (AA) model, i.e.,
\begin{align}\label{A6}
&\tilde{t}[\psi_{2}(n+2)+\psi_{2}(n-2)]+2\lambda \cos(2\pi\beta n+\phi)\psi_{2}(n)\notag\\
&=\tilde{E}\psi_{2}(n).
\end{align}

According to the Aubry-Andr\'{e} self-duality, when
\begin{align}\label{A7}
\tilde{t}=|\lambda|\ \ \rightarrow t^2= |V_{0}(-E^2+m^2)/(2E)|,
\end{align}
there exist localized-extended transitions.
Due to energy-dependence of $\lambda$, the mobility edges would appear in the flat band system.
The mobility edges are determined by Eq.(\ref{A7}). A similar mechanism of localized-extended transition in flat band lattice model has been also investigated in Ref. \cite{Danieli2015}.

\begin{figure}
\begin{center}
\includegraphics[width=1.0\columnwidth]{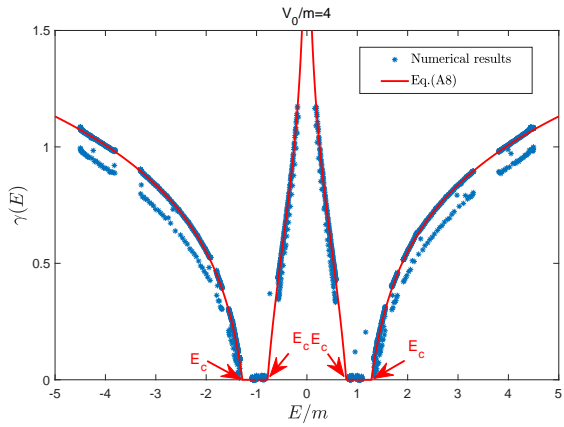}
\end{center}
\caption{ Lyapunov exponent for potential strength $V_0=4m$. The discrete points are numerical results and the solid line is given by Eq.(\ref{A11}).  The mobility edges $E_c\simeq \pm 0.78m, \pm1.28m$ for $V_0=4m$ are indicated by red arrows.  In this whole appendix, we also take $t=m$.}
\label{schematic}
\end{figure}

\begin{figure}
\begin{center}
\includegraphics[width=1.0\columnwidth]{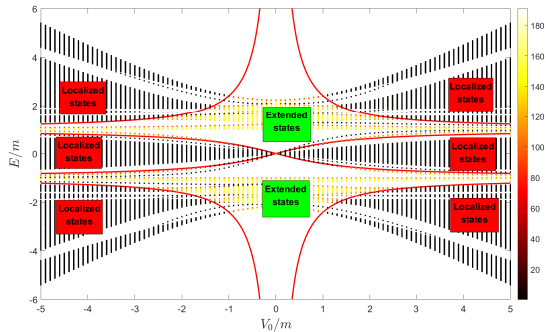}
\end{center}
\caption{ Phase diagram and standard deviations for quasi-periodic potential of type II. The localized state regions and the extended state regions  are labeled in the figure.
 Standard deviations are represented with different colors.}
\label{schematic}
\end{figure}

\begin{figure}
\begin{center}
\includegraphics[width=1.0\columnwidth]{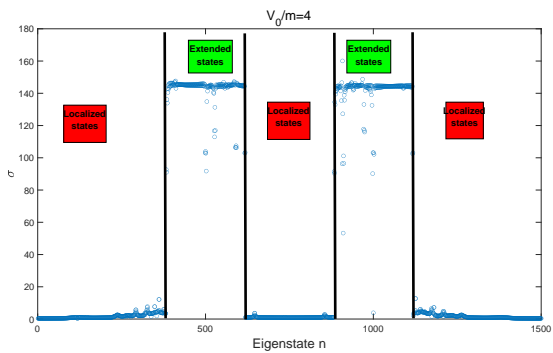}
\end{center}
\caption{ Standard deviations for localized and extended states. The energy  $E$ of states increases as state index $n$ runs from $1$ to $1500$.}
\label{schematic}
\end{figure}

When the parameter $E$ is an eigenenergy,  the Lyapunov exponent for the AA model  is given by \cite{Avila2015}
\begin{align}\label{A11}
&\gamma(E)=\frac{1}{2}Max\{0,\log(|\frac{\lambda}{\tilde{t}}|)\}\notag\\
&=\frac{1}{2}Max\{0,\log(|\frac{V_{0}(E^2-m^2)}{2Et^2}|)\}.
\end{align}
It is noted that  due to two times original lattice constant here [see Eq.(\ref{A6})], there is also  an extra factor $1/2$ in Eq.(\ref{A11}).

Similarly, we also calculate the Lyapunov exponent numerically for  Eq.(\ref{A6}).
 To be specific, taking $t/m=1$, $V_0/m=4$,  we calculate the Lyapunov exponents numerically for all the eigenenergies [see the  set of discrete points in  Fig.6].
 In our numerical calculation, we take $L=125$, phase $\phi=0$, $\psi_2(0)=0$ and $\psi_2(2)=1$.
  The solid line of Fig.6 is given by Eq.(\ref{A11}) with the same parameters. It is shown that most of all discrete points fall onto the solid lines.

We also numerically calculate standard deviations of coordinates of eigenstates [see Eq.(\ref{26}), Figs.7 and 8].
The phase diagram and standard deviations are reported in Fig.7. The standard deviations are represented with different colors in Fig.7. It is shown that in the presence of quasi-periodic potential of type II, the critical region would not appear and  there only exist the localized-extended transitions.
From Fig.8, we can see that for localized states, standard deviations of coordinates are very small. While for extended states, the standard deviations are much larger.

\section*{Acknowledgements}
This work was supported by the NSFC under Grants Nos.
11874127.


\begin{thebibliography}{0}%
\makeatletter
\providecommand \@ifxundefined [1]{%
 \@ifx{#1\undefined}
}%
\providecommand \@ifnum [1]{%
 \ifnum #1\expandafter \@firstoftwo
 \else \expandafter \@secondoftwo
 \fi
}%
\providecommand \@ifx [1]{%
 \ifx #1\expandafter \@firstoftwo
 \else \expandafter \@secondoftwo
 \fi
}%
\providecommand \natexlab [1]{#1}%
\providecommand \enquote  [1]{``#1''}%
\providecommand \bibnamefont  [1]{#1}%
\providecommand \bibfnamefont [1]{#1}%
\providecommand \citenamefont [1]{#1}%
\providecommand \href@noop [0]{\@secondoftwo}%
\providecommand \href [0]{\begingroup \@sanitize@url \@href}%
\providecommand \@href[1]{\@@startlink{#1}\@@href}%
\providecommand \@@href[1]{\endgroup#1\@@endlink}%
\providecommand \@sanitize@url [0]{\catcode `\\12\catcode `\$12\catcode
  `\&12\catcode `\#12\catcode `\^12\catcode `\_12\catcode `\%12\relax}%
\providecommand \@@startlink[1]{}%
\providecommand \@@endlink[0]{}%
\providecommand \url  [0]{\begingroup\@sanitize@url \@url }%
\providecommand \@url [1]{\endgroup\@href {#1}{\urlprefix }}%
\providecommand \urlprefix  [0]{URL }%
\providecommand \Eprint [0]{\href }%
\providecommand \doibase [0]{http://dx.doi.org/}%
\providecommand \selectlanguage [0]{\@gobble}%
\providecommand \bibinfo  [0]{\@secondoftwo}%
\providecommand \bibfield  [0]{\@secondoftwo}%
\providecommand \translation [1]{[#1]}%
\providecommand \BibitemOpen [0]{}%
\providecommand \bibitemStop [0]{}%
\providecommand \bibitemNoStop [0]{.\EOS\space}%
\providecommand \EOS [0]{\spacefactor3000\relax}%
\providecommand \BibitemShut  [1]{\csname bibitem#1\endcsname}%
\let\auto@bib@innerbib\@empty
\end{thebibliography}%


\begin{thebibliography}{10}
\bibitem{Sutherland1986} Bill Sutherland, Localization of electronic wave functions due to local topology, Phys. Rev. B \textbf{34}, 5208 (1986).
\bibitem{Vidal1998} Julien Vidal, R\'{e}my Mosseri, and Benoit Doucot, Aharonov-Bohm Cages in Two-Dimensional Structures, Phys. Rev. Lett. \textbf{81}, 5888 (1998).
\bibitem{Mukherjee} Sebabrata Mukherjee, et al., Observation of a Localized Flat-Band State in a Photonic Lieb Lattice, Phys. Rev. Lett. \textbf{114}, 245504 (2015).


\bibitem{Mielke1999} Andreas Mielke, Ferromagnetism in Single-Band Hubbard Models with a Partially Flat Band, Phys. Rev. Lett. \textbf{82}, 4312 (1999).
\bibitem{Zhang2010} Shizhong Zhang, Hsiang-hsuan Hung, and Congjun Wu, Proposed realization of itinerant ferromagnetism in optical lattices, Phys. Rev.  A \textbf{82}, 053618 (2010).

\bibitem{Shen2010} R. Shen, L. B. Shao, Baigeng Wang, and D. Y. Xing, Single Dirac cone with a flat band touching on line-centered-square optical lattices, Phys. Rev. B \textbf{81}, 041410 (2010).

\bibitem{Urban2011} Daniel F. Urban, Dario Bercioux, Michael Wimmer, Wolfgang H\"{a}usler, Barrier transmission of Dirac-like pseudospin-one particles, Phys. Rev. B \textbf{84}, 115136 (2011).


\bibitem{Fang2016} A. Fang, Z. Q. Zhang, Steven G. Louie, and C. T. Chan, Klein tunneling and supercollimation of pseudospin-1 electromagnetic waves, Phys. Rev. B \textbf{93}, 035422 (2016).


\bibitem{Ocampo2017} Y. Betancur-Ocampo, G. Cordourier-Maruri, V. Gupta, and R. de Coss, Super-Klein tunneling of massive pseudospin-one particles, Phys. Rev. B \textbf{96}, 024304 (2017).

 \bibitem{Tovmasyan2018} Murad Tovmasyan, Sebastiano Peotta, Long Liang, P\"{a}ivi T\"{o}rm\"{a}, and Sebastian D. Huber, Preformed pairs in flat Bloch bands
Phys. Rev. B \textbf{98}, 134513 (2018).

\bibitem{Volovik2019} Volovik, G.E. Flat Band and Planckian Metal. Jetp Lett. \textbf{110}, 352-353 (2019).


\bibitem{Kopnin2011} N. B. Kopnin, T. T. Heikkila, and G. E. Volovik
, High-temperature surface superconductivity in topological flat-band systems, Phys. Rev. B \textbf{83}, 220503(R) (2011).
\bibitem{Iglovikov2014} V. I. Iglovikov, et.al., Superconducting transitions in flat-band systems, Phys. Rev. B \textbf{90}, 094506 (2014).
\bibitem{Peotta2015} S. Peotta,  and  P. T\"{o}rm\"{a},  Superfluidity in topologically nontrivial flat bands. Nat.Commun. \textbf{6}, 8944 (2015).


\bibitem{Julku2016} Aleksi Julku, et. al.,  Geometric Origin of Superfluidity in the Lieb-Lattice Flat Band, Phys. Rev. Lett. \textbf{117}, 045303 (2016).

 \bibitem{Liang2017} Long Liang, et.al.,  Band geometry, Berry curvature, and superfluid weight, Phys. Rev. B \textbf{95}, 024515 (2017).
\bibitem{Cao2018} Yuan Cao, et.al., Unconventional superconductivity in magic-angle graphene superlattices, Nature \textbf{556}, 43 (2018).
  \bibitem{Iskin2019} Iskin, M. Origin of fat-band superfuidity on the Mielke checkerboard lattice. Phys. Rev. A \textbf{99}, 053608 (2019).



\bibitem{Hazra2019} Tamaghna Hazra, Nishchhal Verma, and Mohit Randeria, Bounds on the Superconducting Transition Temperature: Applications to Twisted Bilayer Graphene and Cold Atoms, Phys. Rev. X \textbf{9}, 031049 (2019).

\bibitem{Julku2020} A. Julku, T. J. Peltonen, L. Liang, T. T. Heikkil\"{a}, and P. T\"{o}rm\"{a}, Superfluid weight and Berezinskii-Kosterlitz-Thouless transition temperature of twisted bilayer graphene, Phys. Rev. B  \textbf{101}, 060505(R) (2020).
 \bibitem{Wuyurong2021} Yu-Rong Wu and Yi-Cai Zhang, Superfluid states in $\alpha-T_3$ lattice, Chinese Phys. B \textbf{30}, 060306 (2021).


\bibitem{Wu2021} Yu-Rong Wu, Xiao-Fei Zhang, Chao-Fei Liu, Wu-Ming Liu and Yi-Cai Zhang, Superfluid density and collective modes of fermion superfluid in dice lattice, Sci Rep \textbf{11}, 13572 (2021)



\bibitem{Zhangyicai2021} Yi-Cai Zhang and Guo-Bao Zhu, 2022, Infinite bound states and hydrogen atom-like energy spectrum induced by a flat band, J. Phys. B: At. Mol. Opt. Phys. \textbf{55} 065001. \url{https://iopscience.iop.org/article/10.1088/1361-6455/ac5582/meta}.

\bibitem{Gorbar2019} E. V. Gorbar, V. P. Gusynin, and D. O. Oriekhov, Electron states for gapped pseudospin-1 fermions in the field of a charged impurity, Phys. Rev. B \textbf{99}, 155124 (2019).

\bibitem{Pottelberge2020} R. Van Pottelberge, Comment on ``Electron states for gapped pseudospin-1 fermions in the field of a charged impurity", Phys. Rev. B \textbf{101}, 197102 (2020).
\bibitem{Han2019} Chen-Di Han, Hong-Ya Xu, Danhong Huang, and Ying-Cheng Lai, Atomic collapse in pseudospin-1 systems,  Phys. Rev. B \textbf{99}, 245413 (2019).
\bibitem{Zhangyicai20212}Yi-Cai Zhang 2021, Wave function collapses and 1/n energy spectrum induced by a Coulomb potential in a one-dimensional flat band system, Chinese Phys. B, in press \url{https://doi.org/10.1088/1674-1056/ac3653}.

\bibitem{Zhangyicai20213} Yi-Cai Zhang, Infinite bound states and 1/n energy spectrum induced by a Coulomb potential of
type III in a flat band system, Phys. Scr. \textbf{97}, 015401 (2022) \url{https://iopscience.iop.org/article/10.1088/1402-4896/ac46f4/meta}.

\bibitem{Zhangyicai20214} Yi-Cai Zhang, 2021, Bound States in the Continuum (BIC) Protected By Self-Sustained Potential Barriers in a Flat Band System, \url{https://www.researchgate.net/publication/356223517}.


\bibitem{Sarma1988} S. Das Sarma, Song He, and X. C. Xie, Mobility Edge in a Model One-Dimensional Potential, Phys. Rev. Lett. \textbf{61}, 2144 (1988).
\bibitem{Liu2017} Tong Liu, Gao Xianlong, Shihua Chen, Hao Guo, Localization and mobility edges in the off-diagonal quasiperiodic model with slowly varying potentials, Physics Letters A \textbf{381} (2017) 3683-3687.

\bibitem{Sil2008} Shreekantha Sil, Santanu K. Maiti, and Arunava Chakrabarti, Metal-Insulator Transition in an Aperiodic Ladder Network: An Exact Result, Phys. Rev. Lett. \textbf{101}, 076803 (2008).

\bibitem{Tang2021} Qiyun Tang and Yan He, Mobility edges in one-dimensional models with quasi-periodic disorder, 2021 J. Phys.: Condens. Matter \textbf{33}, 185505.
\bibitem{Sarma1990}  S. Das Sarma, Song He, and X.C. Xie, Localization, mobility edges, and metal-insulator transition in a class
of one-dimensional slowly varying deterministic potentials, Phys. Rev. B \textbf{41}, 5544 (1990).

\bibitem{Duthie2021} Alexander Duthie, Sthitadhi Roy, and David E. Logan, Self-consistent theory of mobility edges in quasiperiodic chains, Phys. Rev. B \textbf{103}, L060201 (2021).
\bibitem{Biddle2010} J. Biddle and S. Das Sarma, Predicted Mobility Edges in One-Dimensional Incommensurate Optical Lattices: An Exactly Solvable Model of Anderson Localization, Phys. Rev. Lett. \textbf{104}, 070601 (2010).

\bibitem{An2021} Fangzhao Alex An, Karmela Padavi, Eric J. Meier, Suraj Hegde, Sriram Ganeshan,
J. H. Pixley, Smitha Vishveshwara, and Bryce Gadway, Interactions and Mobility Edges: Observing the Generalized Aubry-Andr\'{e} Model, Phys. Rev. Lett. \textbf{126}, 040603 (2021).


\bibitem{Liu2021} Yucheng Wang, Xu Xia, Yongjian Wang, Zuohuan Zheng, and Xiong-Jun Liu, Duality between two generalized Aubry-Andr\'{e} models with exact mobility edges, Phys. Rev. B \textbf{103}, 174205 (2021).
\bibitem{Longhi2019} Stefano Longhi, Metal-insulator phase transition in a non-Hermitian Aubry-Andr\'{e}-Harper model, Phys. Rev. B \textbf{100}, 125157 (2019).
\bibitem{Aubry1980}  S. Aubry and G. Andr¡äe, Ann. Israel Phys. Soc \textbf{3}, 18
(1980).

\bibitem{Boers2007}  Dave J. Boers, Benjamin Goedeke, Dennis Hinrichs, and Martin Holthaus, Mobility edges in bichromatic optical lattices, Phys. Rev. A \textbf{75}, 063404 (2007).
\bibitem{Li2017}  Xiao Li, Xiaopeng Li, and S. Das Sarma, Mobility edges in one-dimensional bichromatic incommensurate potentials, Phys. Rev. B \textbf{96}, 085119 (2017).

\bibitem{Izrailev1999}  F. M. Izrailev and A. A. Krokhin, Localization and the Mobility Edge in One-Dimensional Potentials with Correlated Disorder
, Phys. Rev. Lett. \textbf{82}, 4062 (1999).

\bibitem{Luschen2018}  Henrik P. L\"{u}schen,  Sebastian Scherg,  Thomas Kohlert, Michael Schreiber,
Pranjal Bordia, Xiao Li,  S. Das Sarma, and Immanuel Bloch,
Single-Particle Mobility Edge in a One-Dimensional Quasiperiodic Optical Lattice, Phys. Rev. Lett. \textbf{120}, 160404 (2018).

\bibitem{Delyon1984}  Francois Delyon,  Barry Simon, and Bernard Souillard, From Power-Localized to Extended States in a Class
of One-Dimensional Disordered Systems, Phys. Rev. Lett. \textbf{11}, 2187 (1984).
\bibitem{Ganeshan2015} Sriram Ganeshan, J. H. Pixley, and S. Das Sarma, Nearest Neighbor Tight Binding Models with an Exact Mobility Edge in One Dimension, Phys. Rev. Lett. \textbf{114}, 146601 (2015).






\bibitem{Wangyucheng2020} Yucheng Wang, Xu Xia, Long Zhang,  Hepeng Yao, Shu Chen, Jiangong You,
Qi Zhou, and Xiong-Jun Liu, One-Dimensional Quasiperiodic Mosaic Lattice with Exact Mobility Edges, Phys. Rev. Lett. \textbf{125}, 196604 (2020).


\bibitem{YONGJIAN2} Yongjia Wang, Xu Xia, Jiangong You, Zuohuan Zheng, and Qi Zhou, Exact mibility edges for 1D quasiperiodic models, arXiv:2110.00962v1.

\bibitem{Avila2015} Artur Avila, Global theory of one-frequency
Schr\"{o}dinger operators, Acta Math., \textbf{215} (2015), 1-54.




\bibitem{Zhangyicai2022} Yi-Cai Zhang and Yan-Yang Zhang,  Lyapunov exponent, mobility edges and critical region in the generalized Aubry-Andr\'{e} model with an unbounded quasi-periodic potential, Phys. Rev. B 105, 174206 (2022).







\bibitem{Danieli2015} Carlo Danieli, Joshua D. Bodyfelt, and Sergej Flach, Flat-band engineering of mobility edges, Phys. Rev. B \textbf{91}, 235134 (2015).


























\end{thebibliography}
\end{document}